\documentclass[onecolumn,pre]{revtex4}
\usepackage{graphicx, subfigure}
\usepackage{amssymb, amsmath,amssymb,amsfonts}
\usepackage{amsthm,mathrsfs,amsopn}
\usepackage{dcolumn}
\usepackage{bm}
\usepackage{color}
\usepackage[utf8]{inputenc}

\theoremstyle{plain} 

\begin{document}

\title{Dynamical systems on Hypergraphs.}

\author{ Timoteo Carletti$^1$}
\affiliation{$^1$naXys, Namur Institute for Complex Systems, University of Namur, Belgium}

\author{ Duccio Fanelli$^2$}
\affiliation{$^2$Universit\`{a} degli Studi di Firenze, Dipartimento di Fisica e Astronomia,
CSDC and INFN, via G. Sansone 1, 50019 Sesto Fiorentino, Italy}

\author{ Sara Nicoletti$^{2,3}$}
\affiliation{$^2$Universit\`{a} degli Studi di Firenze, Dipartimento di Fisica e Astronomia,
CSDC and INFN, via G. Sansone 1, 50019 Sesto Fiorentino, Italy}
\affiliation{$^3$Dipartimento di Ingegneria dell'Informazione, Universit\`{a} di Firenze,
Via S. Marta 3, 50139 Florence, Italy}

\begin{abstract}
Networks are a widely used and efficient paradigm to model real-world systems where basic units interact pairwise. Many body interactions are often at play, and cannot be modelled by resorting to  binary exchanges. In this work, we consider a general class of dynamical systems anchored on hypergraphs. Hyperedges of arbitrary size  ideally encircle individual units so as to account for multiple, simultaneous interactions. These latter are mediated by a combinatorial Laplacian, that is here introduced and characterised. The formalism of the Master Stability Function is adapted to the present setting. Turing patterns and the synchronisation of non linear (regular and chaotic) oscillators are studied, for a general class of systems evolving on hypergraphs. The response to externally imposed perturbations bears the imprint of the higher order nature of the interactions.
\end{abstract}

\maketitle

\section{Introduction.}
Network science~\cite{AlbertBarabasi,BLMCH} has proved successful in describing many real-world systems~\cite{Newmanbook,Barabasibook,Latorabook}, which, despite inherent differences, share common structural features. Even more interestingly, dynamical processes {and hosting} networks are indissolubly entangled {with the} ensuing patterns {that} reflect in fact the complex topology of the supports {to which they are anchored}
~\cite{Castellanoreview,arenasreview,BBV2008}.

Networks constitute abstract frameworks, where pairwise interactions among generic agents, represented by nodes, are schematised by edges. Stated simply, two agents are connected if they interact. Hence by their very first definition, networks  encode for binary relationships among units. This descriptive framework is sufficiently accurate in many cases of interest, although several examples exist of systems for which it holds true just as a first order approximation~\cite{BGL,LRS2019}. The relevance of high-orders structures has been indeed emphasised in the {context} of functional brain networks~\cite{petri2014homological,LEFGHVD2016}, in applications to protein interaction networks~\cite{estradaJTB}, to the {study} of ecological communities~\cite{GBMSA} and co-authorship networks~\cite{PPV2017,CarlettiEtAl2020}.

Starting from this observation, higher-order models have been developed so as to capture the many body interactions among interacting units. The most notable  examples are  simplicial complexes~\cite{DVVM,BC,PB} and  hypergraphs~\cite{berge1973graphs,estrada2005complex,GZCN}, non trivial mathematical generalisations of ordinary networks that are currently attracting a lot of interest. The concept of simplicial complexes has been for instance invoked to address problems in epidemic spreading~\cite{BKS2016,IPBL} or synchronisation phenomena~\cite{LCB2020,GdPGLRCFLB2020}. Our work is positioned in the framework of hypergraphs, a domain of investigation which is still in its infancy. In this respect, we mention applications to 
social contagion model~\cite{de2019social,ATM2020}, to the modelling of random walks~\cite{CarlettiEtAl2020} and to the study of synchronisation~\cite{Krawiecki2014,MKJ2020} and diffusion~\cite{ATM2020}.

Hypergraphs constitute indeed a very flexible paradigm: an arbitrary number of agents are allowed to interact, thus extending beyond the limit of binary interactions of conventional network models. On the other hand, hypergraphs define a leap forward as compared to simplicial complexes. In this latter case, in fact, if (say) $3$ agents form a $2$-simplex, also all binary interactions are accounted for. 
On the other hand, agents interacting via a hypergraph do form a hyperedge, a unifying frame which encompasses the many body interactions as a whole. Imagine that  a subgroup of agents organised in the hyperedge, also interact with each other via a distinct channel; this yields a new hyperedge, included in the former. A hypergraph can reproduce, in a proper limit, a simplicial complex and, in this respect, provides a more general {tool} for  addressing many body simultaneous interactions.

Furthermore, the analysis of the models in the framework of hypergraphs turns out to be simpler as compared to their simplicial complexes {homologues}. In these latter settings, the involved formulas get rapidly cumbersome and, for this reason, applications are limited to low dimensional simplexes, i.e. $2$ or $3$-simplex. At variance, one can efficiently handle very large hyperedges and, even more importantly,  heterogenous distribution of hyperedges' sizes, because all the information on the high-order structure of the embedding support are stored in a matrix whose dimension depends only on the number of nodes~\cite{CarlettiEtAl2020,ATM2020}.

Starting on these premises, it is clear that many body interactions constitute a relevant and transversal research field that is still in its embryonic stage, in particular as concerns studies that relate to 
hypergraphs. Indeed, novel light could be shed on a large plethora of systems, usually defined on standard networks, by accounting for generalised hypergraph architectures. This paper aims at taking one first step in this direction, by expanding along different axis. We will begin by adapting to the hypergraph setting the Master Stability Function~\cite{Pecora} formalism. We will then consider the
condition for the emergence of Turing patterns~\cite{Turing} for reaction-diffusion systems on hypergraphs, the synchronisation of nonlinear oscillators~\cite{arenasreview} and of chaotic orbits.
It is here anticipated that for theoretical progress to be made one needs to characterise the spectral properties of a properly defined operator, which implements diffusion on hypergraphs.

The Master Stability Function (MSF), is a powerful technique developed in~\cite{Pecora} to analyse synchronisation and it basically amounts to performing a linear stability analysis around a given equilibrium orbit, for a system of coupled interacting units.  A straightforward application of linear stability analysis is for instance found in the context of the celebrated Turing instability, once the reference orbit is indeed a homogeneous fixed point. 

In his seminal paper~\cite{Turing}, Alan Turing set the mathematical basis of pattern formation. Initially proposed to explain the richness and diversity of forms displayed in Nature, the theory elaborated by Turing is nowadays an universally accepted paradigm of self-organisation~\cite{Ball1999,Nicolis1977,Murray2001}. The onset of pattern originates from the loss of stability of an homogeneous equilibrium, as triggered by diffusion. Turing instabilities have been initially studied for systems defined on continuous spatial domains and regular lattices~\cite{OS1971}. More recently, the realm of application of Turing ideas
has been extended to account for reaction-diffusion dynamics hosted on a complex network~\cite{NM2010} and other related structures, such as multilayer networks~\cite{Asllani2014,KHDG} or multigraphs~\cite{Asllani2016} just to mention a few. It is hence a  natural question to generalise these studies to the broad framework of hypergraphs.

Turing patterns emerge from the destabilisation of a homogeneous equilibrium, that is a stationary solution of the examined model. In many real cases, however the system is not bound to evolve close to a stationary solution, but instead displays  periodic oscillations. Examples ranges from biology to ecology, passing through physics~\cite{Pikovsky2001,arenasreview}: individual nonlinear oscillators can  synchronise and thus exhibit a coherent collective behaviour. Synchronisation, the spontaneous ability of coupled oscillators to operate in unison, has been studied for systems {interacting via} a complex and heterogeneous network of interlaced connections. To the best of our knowledge, however, this analysis has never been attempted for systems defined on hypergraphs of the type here considered. Let us observe that, although similar in their {conception}, the works~\cite{Sorrentino2012,BR2014} deal with hypernetworks, namely a network where several different links can connect two nodes, also called multigraph in the literature. {The interactions are hence pairwise}.

The formalism of the MSF can be also applied to chaotic oscillators. The synchronisation of chaotic systems defined on hypergraphs has been studied in~\cite{Krawiecki2014} by using the formalism of the MSF under two main assumptions: (i)  the work has been limited to $p$-hypergraphs, namely assuming that all the hyperedges have the same size $p$; (ii)  the coupling function was assumed to be invariant with respect to permutations of the nodes, within each hyperedge. In this paper, we will relax both assumptions to deal with general hypergraphs with heterogenous hyperedge size distribution and without putting forward any hypothesis on the form of the coupling function.

In a recent work~\cite{MKJ2020}, the synchronisation phenomenon has been studied {resorting again to the} MSF, but employing however a Laplace operator~\cite{JM2019} which {cannot account in full for the high} order interaction {at play}. {The employed operator is defined from the} hyper-adjacency matrix, which is solely capable to encode for the number of incident hyperedges {without gauging their sizes}. Moreover {authors} assumed the coupling function to depend on the average {(arithmetic or geometric)} value of the involved variables. Again, both assumptions are relaxed in the present work, because our Laplace operator  takes into account both the number of incident hyperedges but also their size. {We will moreover make use of} a generic coupling function.


The paper is organised as follows. We first review the formalism of hypergraphs and introduce a new combinatorial Laplace matrix for hypergraphs. We then turn to discussing the spectra of the newly introduced Laplacian by emphasising its localisation properties. Then we present three applications, following the logic path outlined above,  and elaborate on the impact of high-order structures. We finally conclude and sum up of our results.

\section{Hypergraphs.}
\label{sec:hyperg}
Let us consider an hypergraph $\mathcal H(V,E)$, where $V=\{v_1,\dots,v_n\}$ denotes the set of $n$ nodes and $E=\{E_1,\dots,E_m\}$ the set of $m$ hyperedges, that is for all $\alpha=1,\dots,m$: $E_i\subset V$, i.e. an unordered collections of vertices. Note that if $E_\alpha=\{u,v\}$, i.e. $|E_\alpha|=2$, then the hyperedge is actually a ``standard'' edge denoting a binary interaction among $u$ and $v$. If all hyperedges have size $2$ then the hypergraph is actually a network. If an hyperedge contains all its subsets, then we recover a simplicial complex.

\begin{figure}[ht]
\centering
\includegraphics[scale=0.17]{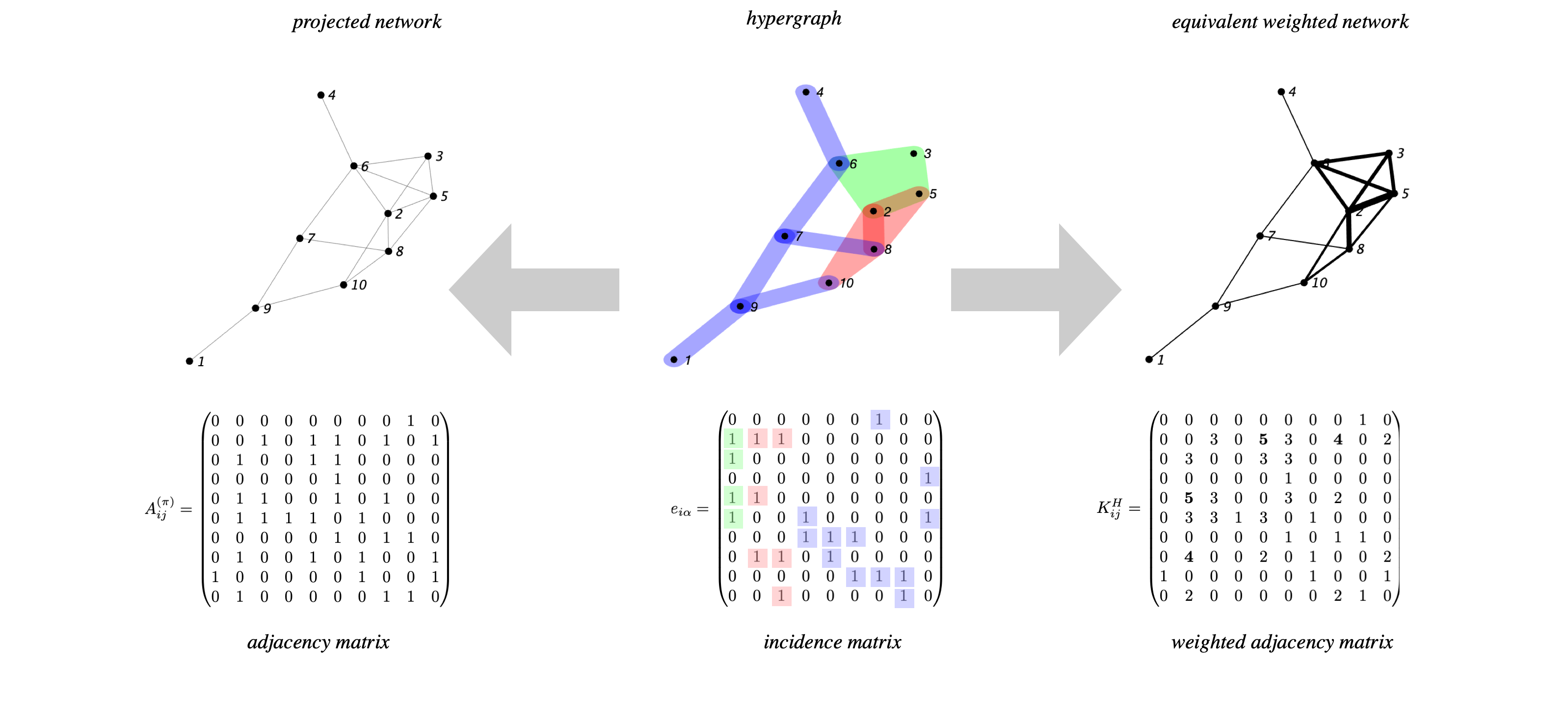}
\caption{\textbf{Hypergraph and networks}. In the middle panel, a  hypergraph is displayed. Hyperedges are coloured according to their size (blue for size $2$, red for size $3$ and green for size $4$). The hypergraph's characteristics are encoded in the incidence matrix $e_{i\alpha}$. Here, the information on how nodes are shared among hyperedges is stored. For ease of visualisation, we coloured the entries of $e_{i\alpha}$ by using the same colour-code that was used to highlight the size of the hyperedges. From the hypergraph, we can construct the projected network, specified by the adjacency matrix $A_{ij}^{(\pi)}$ (left panel), where nodes belonging to the same hyperedge form a complete clique of the suitable size. Alternatively, one can construct the equivalent weighted network (right panel) where the links of the cliques of the projected network are now weighted according to the entries of matrix $\mathbf{K}^H$ (the thicker the line the stronger the weight of the link). The link $(25)$ belongs to a hyperedge of size $3$ and to another one of size $4$. It is therefore the most important of the collection and because of this it receives the largest weight, $k^H_{25}=5$. Observe also the link $(28)$: it belongs to two hyperedges of size $3$ and it is assigned a weight $k^H_{28}=4$, 
larger than the one associated to the links that insist on the hyperedge of size $4$.}
\label{fig:hypergraph}
\end{figure}

We can define the {\em incidence matrix of the hypergraph}~\footnote{We will adopt the convention of using roman indexes for nodes and greek ones for edges.}, $e_{i \alpha}$, which carries information on how nodes are shared among edges (see middle panel Fig.~\ref{fig:hypergraph}). More precisely
\begin{equation}
\label{eq:incid}
e_{i \alpha}=\begin{cases} 1 &\text{$v_i\in E_{\alpha}$}\\
0 & \text{otherwise}\, .
\end{cases}
\end{equation}

With such a matrix one can construct the $n\times n$ adjacency matrix of the hypergraph, $\mathbf{A}=ee^{T}$, whose entry $A_{ij}$ represents the number of hyperedges containing both nodes $i$ and $j$. Note that often the adjacency matrix is defined by setting to $0$ the main diagonal. Let us also define the $m\times m$ hyperedges matrix $\mathbf{C}=e^{T}e$, whose entry $C_{\alpha \beta}$ counts the number of nodes in $E_{\alpha}\cap E_{\beta}$.

The adjacency matrix of the hypergraph allows one to define a Laplace matrix~\cite{JM2019,MKJ2020}, whose entries are given by $k_i\delta_{ij}-A_{ij}$, where $k_i=\sum_j A_{ij}$ denotes the number of hyperedges incident with node $i$. This matrix generalises the (combinatorial) Laplace matrix for networks. However it does not account in full for the higher-order structures encoded in the hypergraph. Notably, the sizes of the incident hyperedges are neglected. 

To overcome this limitation, authors of~\cite{CarlettiEtAl2020}  studied a random walk process defined on a generic hypergraph using a new (random walk) Laplace matrix. It is worth mentioning that the transition rates of the associated process, linearly correlates with the size of the involved hyperedges. Stated differently, exchanges are favoured among nodes belonging to the same hyperedge (weighted according to its associated size). This allows in turn to describe the tightness of high-order interactions among ``close nodes''. More precisely: 
\begin{equation*}
{L}^{RW}_{ij}=\delta_{ij}-\frac{{k}_{ij}^H}{\sum_{\ell\neq i}{k}_{i\ell}^H}\, ,
\end{equation*}
where the entries of $\mathbf{K}^H$ are given by
\begin{equation}
\label{eq:khij}
k^{H}_{ij}=\sum_{\alpha}(C_{\alpha \alpha}-1)e_{i\alpha}e_{j\alpha}=(e\hat{C}e^T)_{ij}-A_{ij}\quad\forall i\neq j \text{ , $k^H_{ii}=0$}\, ,
\end{equation}
and $\hat{C}$ is a matrix whose diagonal coincides with that of $C$ and it is zero otherwise. 

From this random walk Laplace operator, one can straightforwardly derive the (combinatorial) Laplace matrix, 
\begin{equation}
\label{eq:Laphg}
\mathbf{L}^H=\mathbf{D}-\mathbf{K}^H\, ,
\end{equation}
that will be employed in this paper to investigate the effect of diffusion on higher-order structures. In the above equation, matrix $\mathbf{D}$ contains on the diagonal the values $k^H_i=\sum_{{\ell\neq i}} k^H_{i\ell}$ and zeros otherwise. It is clear from its very definition that $\mathbf{K}^H$ takes into account both the number and the size of the hyperedges incident with the nodes. It can also be noted that $\mathbf{K}^H$ can be considered as a {\em weighted adjacency matrix} whose weights have been self-consistently defined to account {for} 
the higher-order structures encoded in the hypergraph (see right panel of Fig.~\ref{fig:hypergraph}).

It is worth emphasising that the dynamics defined on this weighted network is equivalent~\cite{chitraraphael2019} to the dynamics on the corresponding hypergraph. This observation allows us to transport existing tools targeted to networks' analysis to the realm where nodes are made to interact via hypergraphs. In particular, studying linear dynamical systems evolving on a hypergraph amounts to operating with standard $n\times n$ matrices, where $n$ stands for the number of nodes. In this respect, the analysis is straightforward, and avoid the complications  that are to be  faced when dealing with simplicial complexes, where  tensors are instead involved (see Section~\ref{sec:DynHG}).

Given a hypergraph one can construct the {\em projected network}, that is the network obtained by mapping the nodes belonging to a hyperedge into a clique of suitable size (see left panel Fig.~\ref{fig:hypergraph}). If the hypergraph contains only simple hyperedges, then this projection is invertible and, given a network, one can construct a unique hypergraph whose projection coincides with the network itself~\cite{CarlettiEtAl2020}. Let us observe that the projected network keeps track of the many body interactions only though the cliques, i.e. relying on binary exchanges.

Let us conclude this section by remarking that the operator $\mathbf{L}^H$, given by Eq.~\eqref{eq:Laphg}, admits $(1,\dots,1)^T$ as eigenvector associated to the zero eigenvalue. This latter homogeneous solution can be stable, so resilient, to external perturbation for a system evolving on a hypergraph and subject to nonlinear reaction terms. Instabilities can alternatively develop, depending on the specific explored setting. These issues will be addressed in the following by assuming higher-order interactions encoded by the hypergraph, to link co-evolving populations. Inspecting the stability of this generalised class of reaction-diffusion systems, amounts to studying the spectra of the coupling operator. For this reason we shall begin hereafter by analysing the spectra of a hypergraph Laplacian.

\section{Localisation of eigenvectors}
\label{sec:loceigv}

One can prove~\cite{CarlettiEtAl2020} that $\mathbf{L}^H$ is symmetric, non-negatively defined and its largest eigenvalue equals $0$. Moreover, let $\left(\Lambda_H^\alpha\right)_{1\leq \alpha\leq n}$ be the set of its eigenvalues of $\mathbf{L}^H$, then $\Lambda_H^n\geq \dots \Lambda_H^2>\Lambda_H^1=0$, and its eigenvectors, $\left(\vec{\phi}^\alpha\right)_{1\leq \alpha\leq n}$ form an orthonormal basis, $\vec{\phi}^\alpha \cdot \vec{\phi}^\beta=\delta_{\alpha\, \beta}$. As already observed, $\vec{\phi}^1\propto (1,\dots,1)$. Finally $\mathbf{L}^H$ reduces to the Laplace matrix defined on networks once all the hyperedges have size $2$. {In the following we will denote by $\left(\Lambda^\alpha\right)_{1\leq \alpha\leq n}$ the eigenvalues of the Laplace operator of the projected network, $\mathbf{L}$, and $\left(\vec{\psi}^\alpha\right)_{1\leq \alpha\leq n}$ the associated eigenvectors. Based on the well known properties of $\mathbf{L}$ and assuming the network to be connected, we have $\Lambda^n\geq \dots \Lambda^2>\Lambda^1=0$ and the eigenvectors do form an orthonormal basis.}

\begin{figure}[ht]
\centering
\includegraphics[scale=0.4]{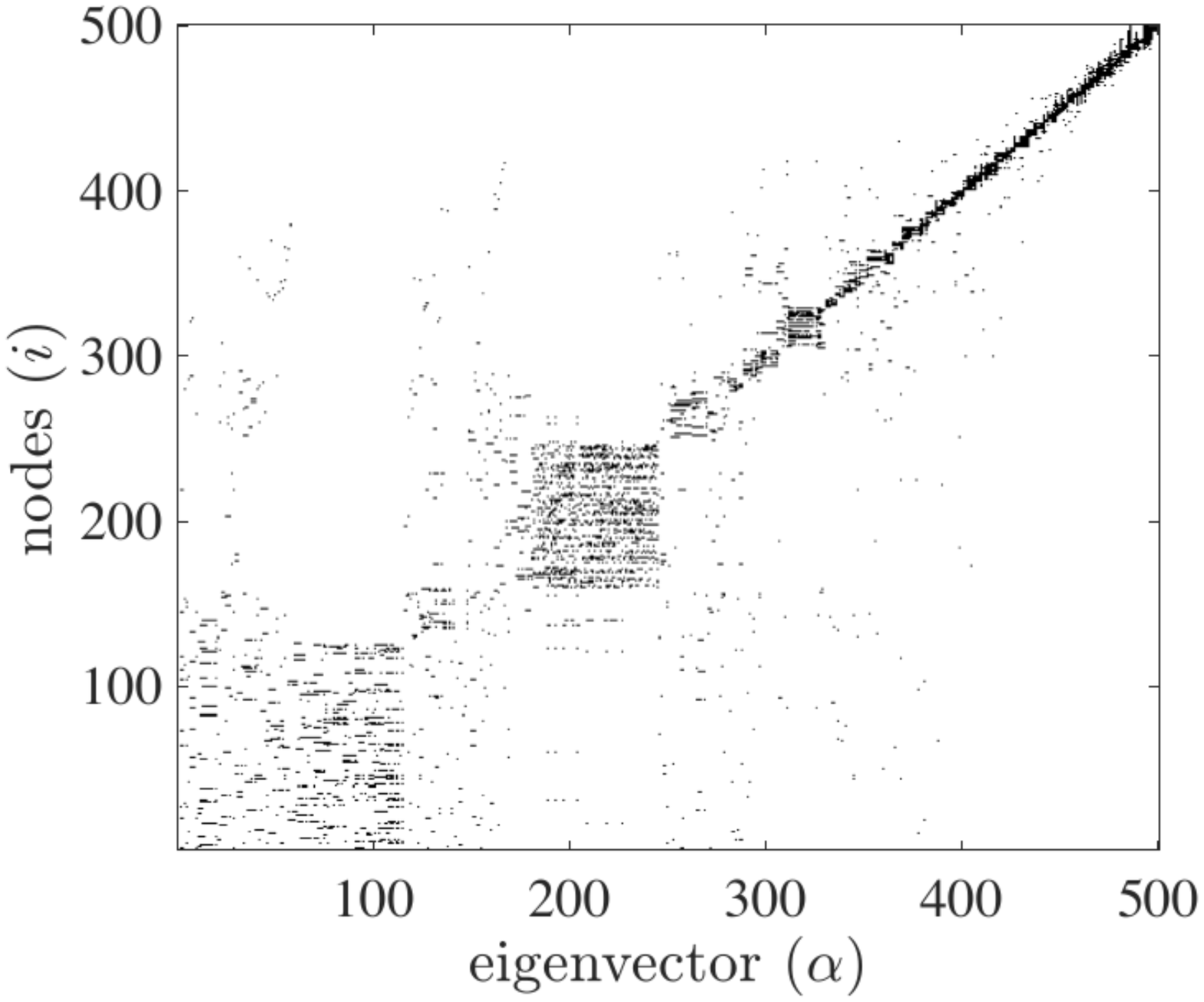}\qquad \includegraphics[scale=0.4]{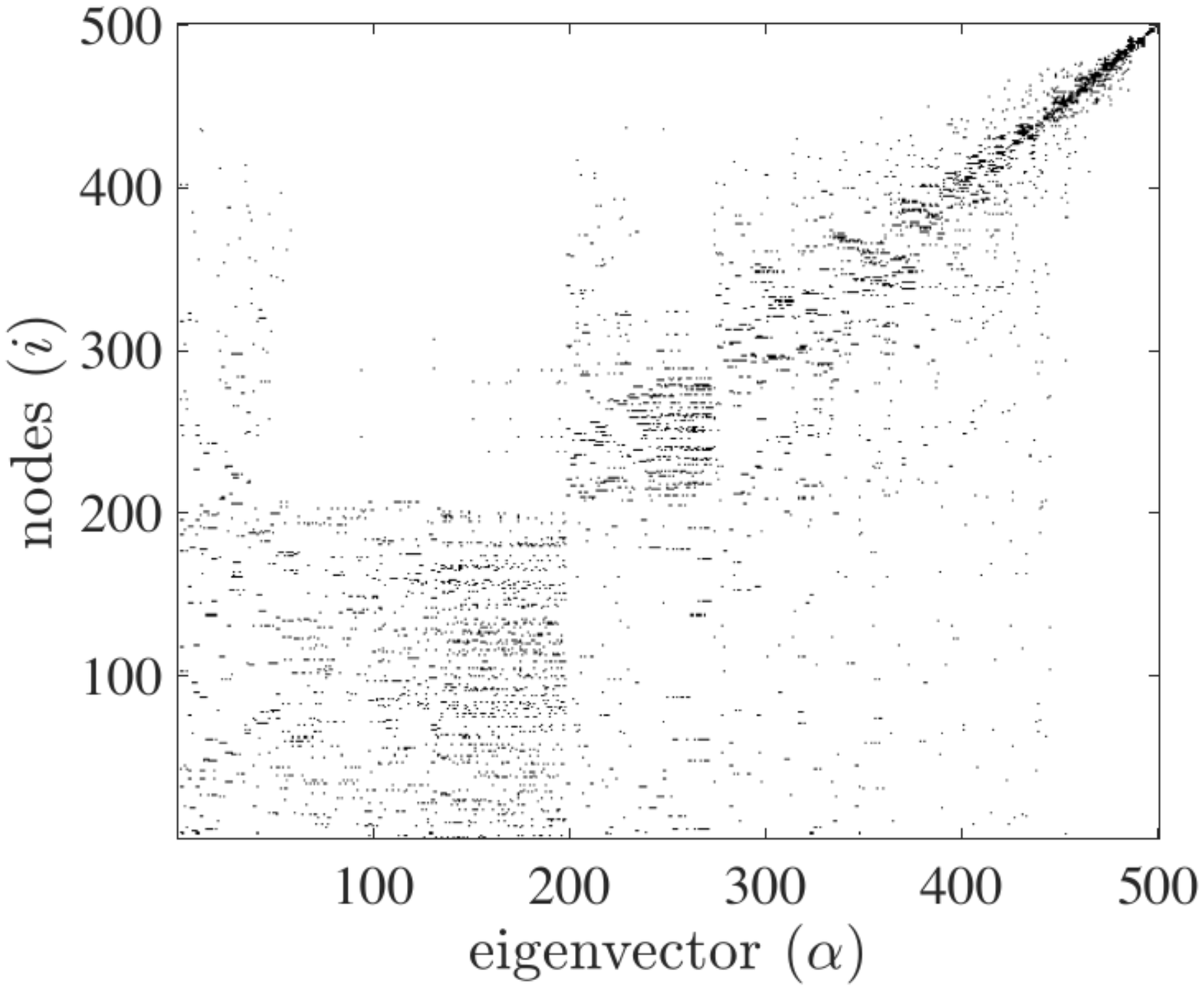}
\vspace{-2cm}
\caption{\textbf{Laplacian eigenvectors}. We report the absolute values of the components of the eigenvectors, $\vec{\phi}^\alpha$, ordered for increasing eigenvalues and nodes degree (right panel) and nodes hyper degree (left panel). Entries larger than $0.015$ are pictured in black, while the remaining ones in white. The projected network is a scale free network made of $500$ nodes and generated by using the configuration method with $\gamma=-2$ and $k_{min}=2$. The corresponding hypergraph is obtained from the latter by transforming all the $m$-cliques into hyperedges of size $m$.}
\label{fig:valpha}
\end{figure}

Localisation of eigenmodes is a phenomenon relevant to many fields of science, e.g. the Anderson localisation in disordered systems~\cite{Anderson1958,GrebenkovNguyen2013}, with a particular relevance to  dynamics. For this reason we decided to start our analysis by studying the localisation properties of the Laplacian eigenvectors for the hypergraph~\eqref{eq:Laphg} and compare them with the corresponding quantities obtained for the projected network. Results reported in Fig.~\ref{fig:valpha} show that  the localisation is more evident for a hypergraph, than for the associated projected network. In the left panel of Fig.~\ref{fig:valpha}, we present the eigenvectors for the Laplace matrix stemming from the hypergraph (ordered for increasing eigenvalue $\Lambda_H^\alpha$) as a function of the nodes indexes (ordered for increasing $k_i^H$). In the right panel, the same quantity is displayed for the Laplace matrix computed from the projected network. In this latter case, the nodes are ordered for increasing degree. By visual inspection (entries larger than $0.015$ are coloured in black while the remaining ones are drawn in white), one can clearly appreciate the dark squarish zones, associated to small or medium rank eigenvectors, which appear in the left panel of Fig.~\ref{fig:valpha}: eigenvectors are found with relatively large entries across many nodes, i.e. a strong localisation. On the right panel, similar structures are present but much weaker. A substantially analogous behaviour is observed for high ranked eigenvectors, e.g. $\alpha \gtrsim 400$ in the left panel and $\alpha \sim 500$ in the right one, for which only few entries display very large values, {pointing hence to} an even stronger localisation (see the thin dark ``line'' in the top right corners in both panels).

\begin{figure}[ht]
\centering
\includegraphics[scale=0.4]{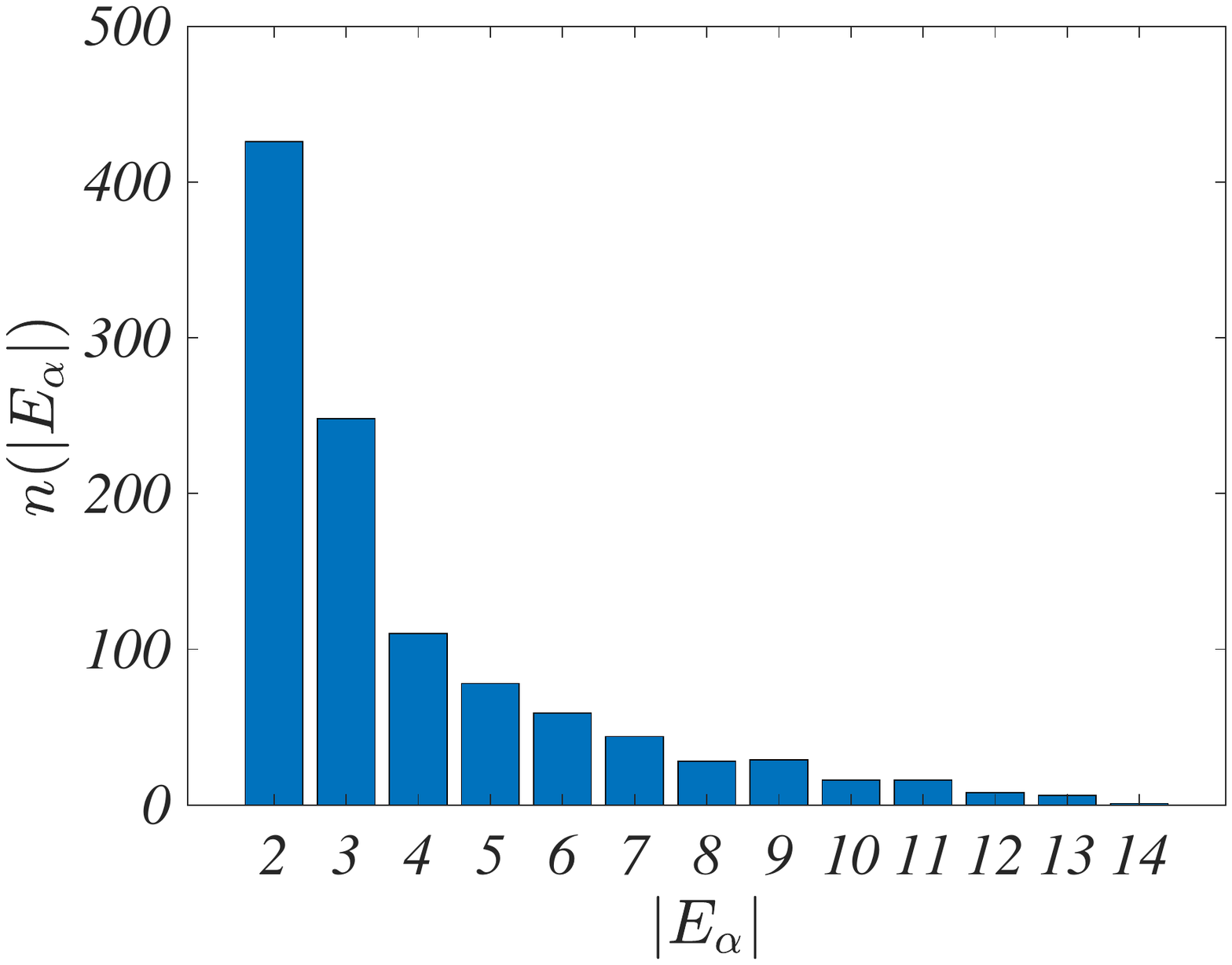}
\vspace{-2cm}
\caption{\textbf{Distribution of hyperedges sizes}. We report the distribution of hyperedges sizes for a hypergraph whose Scale Free projected network is made by $500$ nodes and built using the configuration method with $\gamma=-2$ and $k_{min}=2$. One can observe the presence of relatively large hyperedges responsible for high-order interactions.}
\label{fig:distE}
\end{figure}

To illustrate our results, we employed as projected network a Scale Free network made by $n=500$ nodes, built by using the configuration model with $\gamma=-2$ and $k_{min}=2$~\cite{Latorabook}. The associated hypergraph is obtained by transforming all the maximal $m$-cliques into hyperedges of size $m$. The distribution of hyperedges sizes is reported in Fig.~\ref{fig:distE}.

\begin{figure}[ht]
\centering
\includegraphics[scale=0.4]{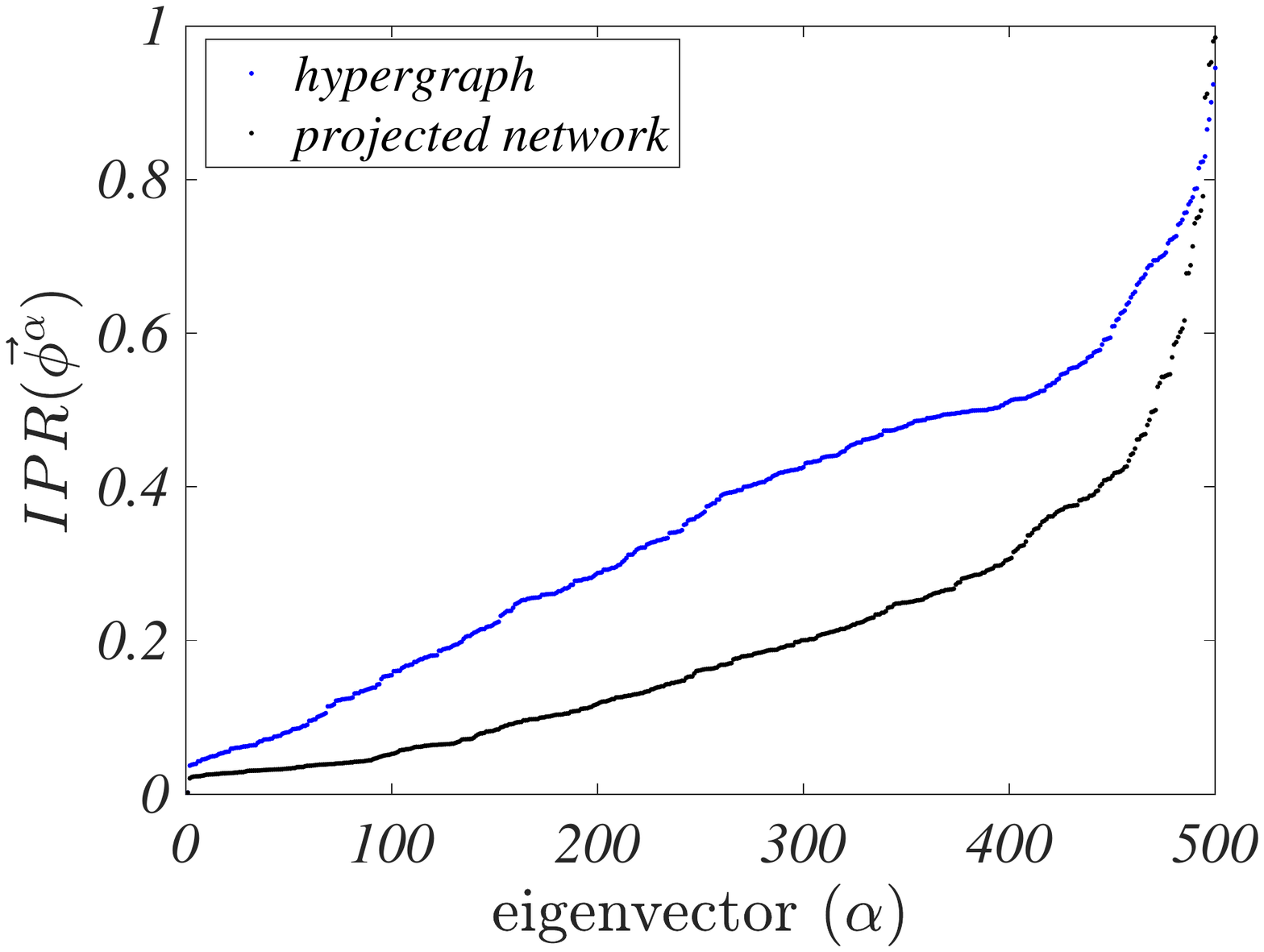}
\vspace{-2cm}
\caption{\textbf{Inverse Participation Ratio}. We report the IPR of the eigenvectors of the hypergraph (blue dots) and of the associated projected network (black dots) used in Fig.~\ref{fig:valpha}. We can observe that in the case of the hypergraph the IPR is always larger than that obtained for the corresponding projected network, while the eigenvectors associated to the largest eigenvalues are more localised for the case of the network.}
\label{fig:IPR}
\end{figure}

A more quantitative measure of the localisation, can be  obtained by using the {\em Inverse Participation Ratio} (IPR)~\cite{McGrawMenzinger}. For a $n$-dimensional vector, $\mathbf{v}$, this is defined as
\begin{equation}
\label{eq:IPR}
P(\mathbf{v})=\frac{\sum_i v_i^4}{\left(\sum_i v_i^2\right)^2}\, .
\end{equation}
The above quantity ranges in $[1/n,1]$, where the lower bound is attained for a vector with uniform entries. The upper limit is hit  when all entries are $0$ but one, which equals $1$.
In Fig.~\ref{fig:IPR} we report the IPR computed for the eigenvectors of the hypergraph (blue dots) and the projected network (black dots) used in Fig.~\ref{fig:valpha}. We can observe that in the case of the hypergraph, the IPR is always larger than the homologous quantity computed for the projected network, except for very high ranked eigenvectors (say, the last $5$ ones). 

In the next section we will show that the localisation which manifests on hypergraph, leaves macroscopic imprints on the  dynamics of systems subject to many-body, higher-order interactions. This issue will be discussed in the following Section.

\section{Dynamical systems on hypergraphs}
\label{sec:DynHG}

In the remaining part of this paper we will consider the behaviour of dynamical systems defined on hypergraphs. In particular, we will analyse the consequences of dealing with higher-order couplings, exploiting to this end the spectral characteristics highlighted above. More specifically, assume $n$ copies of the same low dimensional dynamical system to be hosted on each node of a given collection. This defines the local dynamics of the inspected system. Units belonging to different nodes are assumed to interact through higher-order structures identified as hyperedges. Many body interactions  promote a preferential interaction among nodes belonging to the same large hyperedge. The nodes can be imagined to identify different spatial locations. For this reason we will denote by aspatial the system composed by one isolated dynamical unit, and use spatial to refer to its multi-dimensional version made of mutually entangled  components.

As already mentioned the newly introduced (combinatorial) Laplace matrix~\eqref{eq:Laphg} admits a homogeneous eigenvector associated to the zero eigenvalue. This will allow us to probe (in)stability of interconnected systems evolving close to reference orbits. For the sake of completeness, we will consider three distinct applications that cover several relevant research domains. We will {begin by imposing} a generalised diffusive coupling among nodes as exemplified by the aforementioned Laplace matrix~\eqref{eq:Laphg}. Working in this framework, we will study the emergence of Turing patterns, that is the conditions that promote the emergence of a stable heterogeneous solution. We will then turn to considering the synchronisation between nonlinear oscillators, diffusively coupled via higher-order combinatorial Laplacians. Finally, we will analyse the synchronisation of chaotic oscillators, in the setting of interest where higher-order interactions are at play. The formalism of the Master Stability Function, will be used to tackle the problem analytically. Projected networks will be employed as reference benchmarks to bring into evidence the role of hypergraphs and related higher order interactions. 

Consider a $d$-dimensional system described by local, i.e. aspatial, equations:
\begin{equation}
\label{eq:dotxF}
\frac{d\mathbf{x}}{dt}=\mathbf{F}(\mathbf{x})\quad \mathbf{x}\in\mathbb{R}^d\, ,
\end{equation}
and fix a reference orbit, $\mathbf{s}(t)$. Let us observe that the latter can also be a fixed point. Assume further $n$ identical copies of the above system coupled through a hypergraph, namely each copy is attached to a node of a hypergraph. Moreover, each unit belongs to one (or more) hyperedge. Units sharing the same hyperedge are tightly coupled, due to existing many body interactions. In formulas:
\begin{equation*}
\frac{d\mathbf{x}_i}{dt}=\mathbf{F}(\mathbf{x}_i) -\varepsilon \sum_{\alpha:i\in E_\alpha}\sum_{j\in E_\alpha, j\neq i}(C_{\alpha\,\alpha}-1)\left(\mathbf{G}(\mathbf{x}_i)-\mathbf{G}(\mathbf{x}_j)\right)\, ,
\end{equation*}
where $\mathbf{x}_i$ denotes the state of the $i$-th unit, i.e. anchored to the $i$-th node, $\varepsilon$ the strength of the coupling and $\mathbf{G}$ a generic nonlinear coupling function. 
The elements $C_{\alpha\,\alpha}$ of matrix $\mathbf{C}$ denote the size of the hyperedge $E_\alpha$. The factor $-1$ account for the fact that $j$ should be different from $i$. Recalling the definition of $e_{i\alpha}$ one can rewrite the previous formula as
\begin{eqnarray}
\label{eq:maineq}
\frac{d\mathbf{x}_i}{dt}&=&\mathbf{F}(\mathbf{x}_i) -\varepsilon \sum_{\alpha,j} e_{i\alpha}e_{j\alpha}(C_{\alpha\,\alpha}-1)\left(\mathbf{G}(\mathbf{x}_i)-\mathbf{G}(\mathbf{x}_j)\right)\notag\\
&=&\mathbf{F}(\mathbf{x}_i) -\varepsilon\sum_{j} k_{ij}^H\left(\mathbf{G}(\mathbf{x}_i)-\mathbf{G}(\mathbf{x}_j)\right)=\mathbf{F}(\mathbf{x}_i) -\varepsilon\sum_{j} \left(\delta_{ij}k_{i}^H-k_{ij}^H\right)\mathbf{G}(\mathbf{x}_j)\notag\\
&=&\mathbf{F}(\mathbf{x}_i) -\varepsilon\sum_{j} L^H_{ij}\mathbf{G}(\mathbf{x}_j)\, ,
\end{eqnarray}
where we have used the definition of $k_i^H=\sum_jk_{ij}^H$ and $L^H_{ij}$ given by~\eqref{eq:Laphg}. Let us stress once again that all the high-order structure is encoded in a $n\times n$ matrix and there is no need for tensors as in the case of simplicial complexes: this simplifies the resulting analysis.

By exploiting the fact that $\sum_j L^H_{ij}=0$ for all $i=1,\dots, n$, it is immediate to conclude that the aspatial reference solution $\mathbf{s}(t)$ is also a solution of Eq.~\eqref{eq:maineq}. A natural question that arises is hence to study the stability of the homogeneous solution for the system in its coupled variant. 

To answer to this question one introduce the deviations from the reference orbit, i.e. $\delta\mathbf{x}_i=\mathbf{x}_i-\mathbf{s}$. Assuming this latter to be small, one can derive a self-consistent set of linear equations for tracking the evolution of the perturbation in time. To this end,  we make use of the above expression  in Eq.~\eqref{eq:maineq} and  perform a Taylor expansion by neglecting terms of order larger than two, to eventually get:
\begin{equation}
\label{eq:GLHGlin}
\frac{d\delta\mathbf{x}_i}{dt}=D\mathbf{F}(\mathbf{s}(t))\delta\mathbf{x}_i -\varepsilon \sum_{j} {L}^H_{ij} D\mathbf{G}(\mathbf{s}(t))\delta\mathbf{x}_j\, , 
\end{equation}
where $D\mathbf{F}(\mathbf{s}(t))$ (resp. $D\mathbf{G}(\mathbf{s}(t))$) denotes the Jacobian matrix of the function $\mathbf{F}$ (resp. $\mathbf{G}$) evaluated on the trajectory $\mathbf{s}(t)$.

Remember that $\mathbf{L}^H$ is symmetric. Hence, there exists  a basis formed by orthonormal eigenvectors, $\vec{\phi}^\alpha$, associated to eigenvalues $\Lambda_H^\alpha$ (see Section~\ref{sec:loceigv}). We can then project $\delta\mathbf{x}_i$ on this basis and obtains for all $\alpha$:
\begin{equation}
\label{eq:GLHGlinalpha}
\frac{d\delta\mathbf{y}_\alpha}{dt}=\left[D\mathbf{F}(\mathbf{s}(t))-\varepsilon \Lambda_\alpha D\mathbf{G}(\mathbf{s}(t))\right]\delta\mathbf{y}_\alpha\, , 
\end{equation}
where $\delta\mathbf{y}_\alpha$ is the projection of $\delta\mathbf{x}_i$ on the $\alpha$-th eigendirection. 

Let us finally conclude this section by observing that from Eq.~\eqref{eq:GLHGlinalpha} one can derive the Master Stability Function, i.e. the most general framework  to address questions that pertain to the stability of the reference orbit. Despite its generality, the latter can only be handled numerically, except very few exceptions. {In the following we begin by studying the setting where} $\mathbf{s}(t)$ is a constant solution. In this case the Eq.~\eqref{eq:GLHGlinalpha} simplifies because the right hand side is no longer time dependent and the problem reduces to a classical  study of Turing instability. Indeed, the rightmost term in Eq.~\eqref{eq:maineq} {can be seen} as a sort of generalised Fickean diffusion (see Section~\ref{ssec:TP}). If the reference orbit is instead periodic in time, {one can investigate the conditions which drive} the synchronisation of regular oscillators. {In this case} the Master Stability Function can be analysed by resorting to the Floquet machinery. In the following, we have however chosen to study the synchronisation of Stuart-Landau oscillators via higher-order couplings (see Section~\ref{ssec:GL}). Working in this setting,  the Master Stability Function becomes again time independent and the analysis {closely resembles the one carried out for addressing the onset of Turing instabilities}. {As a final step, we will turn to} studying the case where $\mathbf{s}(t)$ is a chaotic trajectory (see Section~\ref{ssec:MSF}).

\subsection{Turing patterns on hypergraphs}
\label{ssec:TP}
The Turing instability {takes place for} spatially extended systems:  a stable homogeneous equilibrium becomes unstable upon injection of a heterogeneous, i.e. spatially dependent, {perturbation} once diffusion and  reaction terms are simultaneously at play. Let us first consider two generic nonlinear functions $f(u,v)$ and $g(u,v)$ describing the local dynamics
\begin{equation}
\label{eq:locdyn}
\begin{cases}
\dot{u} &=f(u,v)\\
\dot{v} &=g(u,v) \, .
\end{cases}
\end{equation}
Then assume to replicate such system on all the nodes of a given hypergraphs, and label $u_i$ and $v_i$ {the corresponding concentration. Here} the index $i$ refers to the specific node to which the dynamical variables are bound. Finally, assume that two nodes, $i$ and $j$, communicate if they belong to the same hyperedge and moreover the strength of the interaction (which results in an effective transport across the involved nodes) is mediated by both the number of shared hyperedges and their sizes. Indeed, nodes belonging to the same hyperedge exhibit a higher-order interaction and we consequently assume that spreading among them is more probable than with nodes associated with other hyperedges or smaller ones. From a microscopic point of view, imagine to deal with a  walker belonging to a given node. The walker assigns to all its neighbours a weight that gauges the size of the hyperedges and the number of incident hyperedges, and then she performs a jump with a probability proportional to this weight. This represents a higher-order extension Ficks' law: the  rate of change of $u_i$ is proportional to 
\begin{equation*}
\dot{u}_i\sim \sum_{\alpha: i\in E_\alpha}\sum_{j\in E_\alpha, j\neq i} (C_{\alpha\,\alpha}-1)(u_i-u_j)\, ,
\end{equation*}
where use  has been made of matrix $\mathbf{C}$, as introduced above. Recalling the definition of $e_{i\alpha}$ one can rewrite the previous formula as
\begin{equation*}
\dot{u}_i\sim\sum_{\alpha,j} e_{i\alpha}e_{j\alpha}(C_{\alpha\,\alpha}-1)(u_i-u_j)=\sum_{j} k_{ij}^H(u_i-u_j)=\sum_{j} \left(\delta_{ij}k_{i}^H-k_{ij}^H\right)u_j=\sum_j L^H_{ij}u_j\, ,
\end{equation*}
where we have used the definition of $k_i^H=\sum_jk_{ij}^H$ and $L^H_{ij}$.

So in conclusion a reaction-diffusion processes on hypergraphs, where the diffusion takes into account the higher-order interactions among nodes in the same hyperedge, can be described by the following system
\begin{equation}
\label{eq:glodyn}
\begin{cases}
\dot{u_i} &=f(u_i,v_i)+D_u \sum_j L^H_{ij}u_j\\
\dot{v_i} &=g(u_i,v_i) +D_v \sum_j L^H_{ij}v_j
\end{cases}\, ,
\end{equation}
where $D_u$ and $D_v$ are effective diffusion coefficients of species $u$ and $v$. At first sight, the above model seems to solely account for binary interactions. However, higher-order interactions are also present, as encoded in the matrix $\mathbf{L}^H$. This is thus a compact formalism allowing to overcome the computational issues {intrinsic to} simplicial complexes. Finally, let us observe that if the hypergraph is a network, i.e. the hyperedges have size $2$, $|E_{\alpha}|=2$ $\forall \alpha$, then $\mathbf{L}^H$ reduces to the standard Laplace matrix. Thus Eqs. (\ref{eq:glodyn}) converges to the standard reaction-diffusion system defined on a network. 

The {condition for the emergence of a} Turing instability can be detected by performing a linear stability analysis about the homogeneous equilibrium. More precisely, the latter is assumed to be stable with respect to homogeneous perturbations, while it loses its stability for heterogeneous perturbations once diffusion is at play, $D_u>0$ and $D_v>0$. The linear stability analysis can be performed by following the standard procedure: (i) by linearising the model~\eqref{eq:glodyn} around the equilibrium, $(u_i,v_i)=(\bar{u},\bar{v})$ for all $i$; (ii) by expanding the perturbations on the eigenbasis of $\mathbf{L}^H$, and (iii) by calculating the dispersion relation, i.e. the linear growth rate  $\lambda_\alpha = \lambda(\Lambda_H^\alpha)$ of the eigenmode $\alpha$, as a function of the Laplacian eigenvalue $\Lambda_H^\alpha$.
The linear growth rate is the real part of the largest root of the second order equation
\begin{equation}
\label{eq:reldisp}
\lambda_\alpha^2 -\lambda_\alpha\left[ \mathit{tr} \mathbf{J}_0+\Lambda_H^\alpha(D_u+D_v)\right]+\det \mathbf{J}_0+\Lambda_H^\alpha(D_u\partial_v g+D_v\partial_u f)+D_uD_v (\Lambda_H^\alpha)^2=0\, ,
\end{equation}
where $\mathbf{J}_0=\left(
\begin{smallmatrix}
 \partial_u f & \partial_v f\\
  \partial_u g & \partial_v g
\end{smallmatrix}
\right)$ is the Jacobian matrix of the reaction part evaluated at the equilibrium $(u_i,v_i)=(\bar{u},\bar{v})$, $\mathit{tr}$ (resp. $\det$) is its trace (resp. determinant). The concept of dispersion relation is close to that of Lyapunov exponent:  the existence of eigenvalues $\Lambda_H^\alpha$ for which the dispersion relation takes positive values, implies that the system goes unstable via a typical path first identified by  Alan Turing in his seminal work. At variance, if the dispersion relation is  negative the system cannot undergo a Turing instability: any tiny perturbation fades away and the system settles back to the homogeneous equilibrium.

\begin{figure}[ht]
\centering
\includegraphics[scale=0.55]{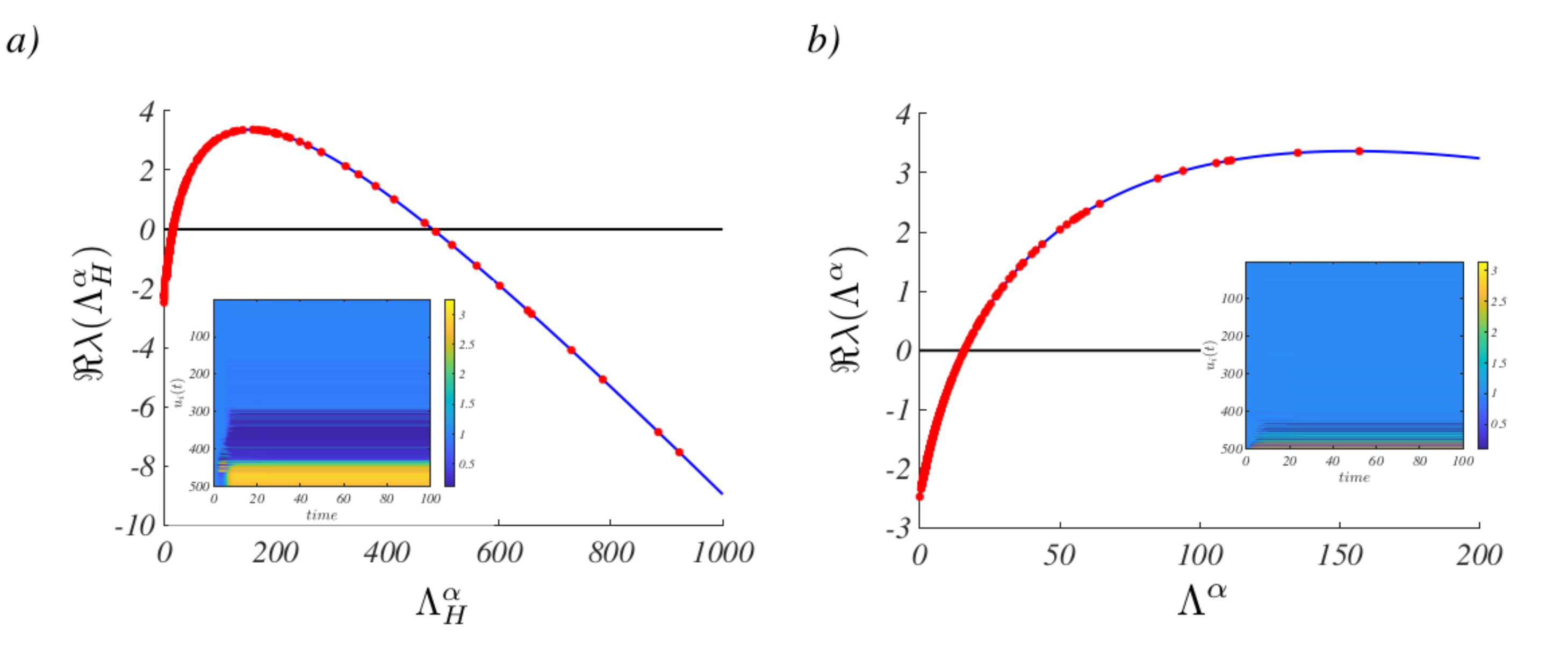}
\caption{\textbf{Turing patterns on hypergraphs}. Main panels: The \textbf{Dispersion relation} for the Brusselator model defined on the hypergraph -- panel (a)-- and the projected network --panel (b)--. One can observe that in both cases there are eigenvalues for which the dispersion relation is positive (red dots); the blue line represents the dispersion relation for the Brusselator model defined on a continuous regular support. Being both Laplace matrices symmetric, the dispersion relation computed for the discrete spectra lies on top of the one obtained for the continuous support. Insets: The \textbf{Turing Patterns} on the hypergraph (panel (a)) and the projected network (panel (b)). We report the time evolution of the concentration of the species $u_i(t)$ in each node as a function of time, by using an appropriate colour code (yellow associated to large values, blue to small ones). In the former case, nodes are ordered for increasing hyper degree while in the second panel for increasing degree. One can hence conclude that nodes associated to large hyper degrees display a large concentration amount for species $u_i$. This yields a very localised pattern. The hypergraph and the projected network are the same used in Fig.~\ref{fig:valpha}.}
\label{fig:reldisp}
\end{figure}

To provide a concrete example, we assume the reaction kinetic to be modelled by the  Brusselator scheme~{\cite{PrigogineNicolis1967,PrigogineLefever1968}}. This is a nonlinear model defined by $f(u,v)=1-(b+1)u+c u^2v$ and $g(u,v)=bu-cu^2v$, where $b$ and $c$ act as tunable parameters. We first show an example of Turing pattern emerging in both the hypergraph and its related projected network (the same used in the previous section). In the main panels of Fig.~\ref{fig:reldisp}  the dispersion relations are reported:  a subset of eigenvalues exist which is associated to positive values of the dispersion relation, for both the hypergraph --panel (a)-- and the projected network --panel (b). In the insets of Fig.~\ref{fig:reldisp} we display the ensuing patterns. Nodes are ordered for increasing hyper degree (resp. degree) for the hypergraph (resp. the projected network).  One can clearly observe that, in the case of the hypergraph, patterns are strongly localised in nodes associated to larger hyper degree.
%

\begin{figure}[ht]
\centering
\includegraphics[scale=0.35]{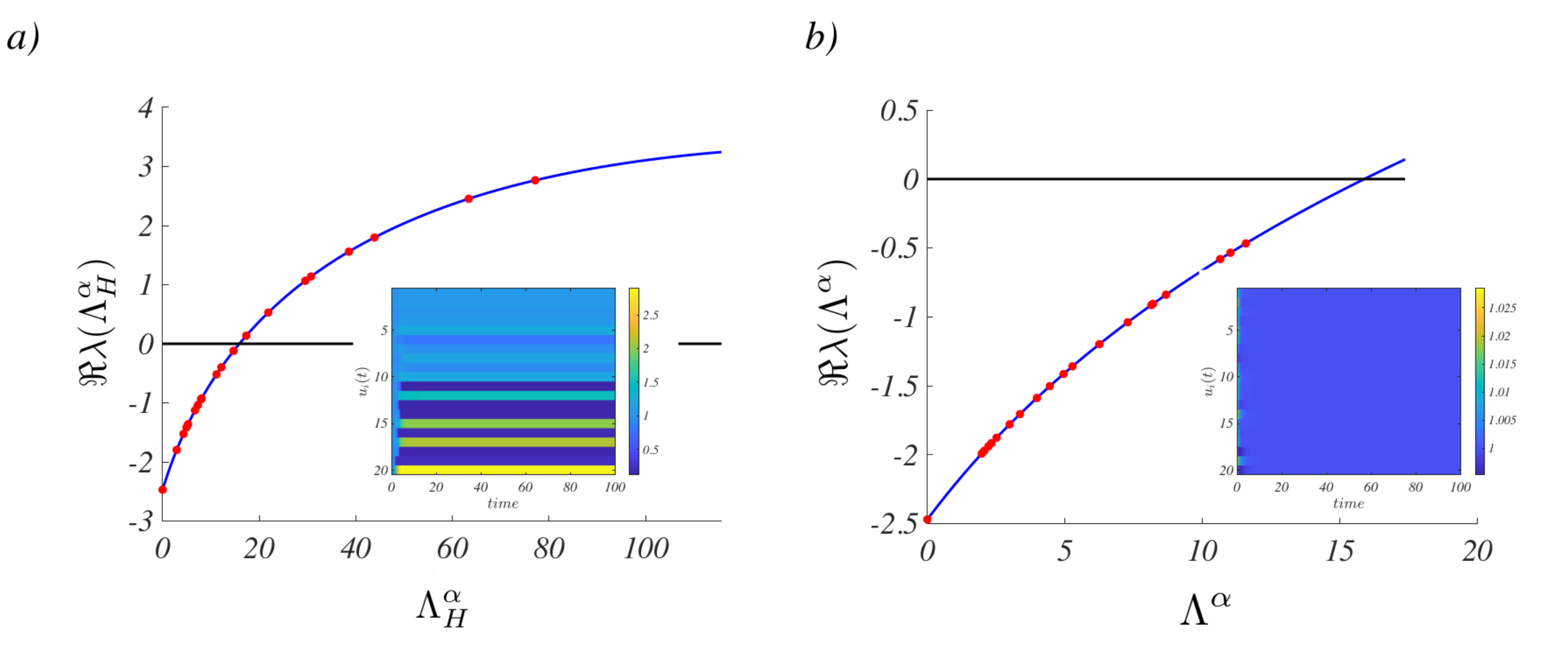}
\caption{\textbf{Many body induced Turing patterns}. Main panels: The \textbf{Dispersion relation} for the Brusselator model defined on the hypergraph (panel (a)) and the projected network (panel (b)). In the former case eigenvalues are found for which the dispersion relation is positive (red dots) while for the system defined on the projected network this conclusion does not hold. The blue line represents again the  dispersion relation for the Brusselator model on a continuous support. Insets: Because the condition for Turing instability is satisfied for the hypergraph Laplace matrix, \textbf{Turing Patterns} emerge on the hypergraph (panel (a)). At variance, patterns do not manifest on the projected network (panel (b)). We report the time evolution of the concentration of species $u_i(t)$, on each node, as a function of time by using a proper colour code (yellow associated to large values, blue to small ones). In the former case nodes are ordered for increasing hyper degree while in the latter for increasing degree. The hypergraph and the projected network are obtained by means of the Barab\'asi-Albert algorithm~\cite{AlbertBarabasi} with $20$ nodes. Every iteration, $3$ newly added nodes are attached to the existing ones.}
\label{fig:reldispb}
\end{figure}

From Fig.~\ref{fig:reldisp} one can also observe that the domain of definition of the eigenvalues for the hypergraph cover a much wider range, as compared to that associated to the projected network. This observation can open the way to settings where patterns emerge only for systems defined on top of hypergraphs and not on the corresponding projected networks. In this case, patterns are the result of the higher-order interaction among nodes. To challenge this scenario, let us  consider a small network built by using the Barab\'asi-Albert algorithm~\cite{AlbertBarabasi} with $20$ nodes. For each iteration of the generative algorithm, $3$ new nodes are attached to the already existing ones, according to a preferential attachment scheme; because of the small size of the network, our goal here is not to resolve the  scale free nature of the network but to obtain a hierarchical structure where $3$-cliques, and larger ones, are mutually connected. We identify the complete cliques and  build the associated hypergraph by assuming each $m$-clique to form a hyperedge with size $m$. We then turn to considering the resulting hypergraph and the associated projected network as the underlying support for the dynamics~\eqref{eq:glodyn}. The dispersion relation can be computed (see main panels of Fig.~\ref{fig:reldispb}): observe that the homogenous equilibrium is stable even in presence of diffusion on the network while it looses stability in the case of the hypergraph. In this latter setting Turing patterns are hence expected to develop. This can be checked by computing the time evolution of the species density $u_i(t)$ both on the hypergraph and the projected network. By inspection of  Fig.~\ref{fig:reldispb} one can appreciate that heterogenous patterns develop in the former case (see inset in the panel (a) of Fig.~\ref{fig:reldispb}). Patterns are instead lacking in the latter scenario, i.e. when the Brussellator model hosted on the projected network (see inset in the panel (b) of Fig.~\ref{fig:reldispb}).

%

\subsection{Synchronisation of Stuart-Landau oscillators on hypergraphs}
\label{ssec:GL}

In the previous section we studied the emergence of Turing patterns for reaction-diffusion systems defined on a hypergraph so as to account for many body interactions. These patterns originate from a symmetry breaking instability induced by an externally imposed perturbation acting on systems initially close to a stationary homogeneous equilibrium. In many relevant problems, systems display  periodic solutions. It is therefore important to investigate the stability of isolated periodic orbits and, even more essential, to study the dynamics of extended systems which combine several replica of the same  nonlinear oscillators. Imagine that individual oscillators are evolving in phase and introduce a non homogeneous perturbation. If the system is globally stable the perturbation gets eventually re-absorbed and the oscillators display a synchronous dynamics~\cite{arenasreview}. Otherwise the perturbation develops in time and the system evolves towards a distinct, heterogeneous, attractor. 

To study the synchronisation  via a hypergraph, we consider individual units obeying to a  Stuart-Landau (SL) equation~\cite{Stuart1978,Kuramoto}. This is a paradigmatic model of nonlinear oscillators, often invoked for modelling a wide range of phenomena, from nonlinear waves to second-order phase transitions, from superconductivity and superfluidity to Bose-Einstein condensation~\cite{AK2002}
Besides, the SL equation can be considered as a normal form for systems close to a supercritical Hopf-bifurcation. In this respect, the results here presented are more general than the specific setting explored. 

Consider an ensemble made of $n$ nonlinear oscillators and label with $W_i$ their associated complex amplitude. Each oscillator obeys a complex Stuart-Landau equation
\begin{equation*}
\frac{d}{dt}W_j=W_j-(1+ic_2)|W_j|^2W_j\, ,
\end{equation*}
where $c_2$ is a real parameter and $i=\sqrt{-1}$. Let us observe that the former admits the limit cycle solution $W_{LC}(t)=e^{-ic_2t}$.

We then assume the oscillators to be coupled via a many body diffusive-like interaction which can be described by the discrete Laplacian matrix~\eqref{eq:Laphg}, returning thus the system
\begin{equation}
\label{eq:GL}
\frac{d}{dt}W_j=W_j-(1+ic_2)|W_j|^2W_j-(1+ic_1)K\sum_kL^H_{jk}W_k\, ,
\end{equation}
where $c_1$ is a second real parameters and $K$ is a suitable parameter setting the coupling strength. Based on the properties of the Laplace matrix, one can prove that the limit cycle solution, $W_{LC}(t)$, is also a solution of Eq.~\eqref{eq:GL}. To characterise the stability of the latter to heterogeneous perturbation, we rewrite $W_j$ using polar coordinates as:
\begin{equation}
\label{eq:W}
W_j(t)=W_{LC}[1+\rho_j(t)]e^{i\theta_j(t)}\, .
\end{equation}
Assuming $|\rho_i(t)|$ and $|\theta_i(t)|$ to be small, one can insert the~\eqref{eq:W} into Eq.~\eqref{eq:GL} and then linearise the resulting equation, to get:
\begin{equation}
\label{eq:vareqGL}
\frac{d}{dt}
\begin{pmatrix}
\rho_j \\
\theta_j
\end{pmatrix}
=
\begin{pmatrix}
-2 & 0 \\
-2c_2 & 0
\end{pmatrix}
\begin{pmatrix}
\rho_j \\
\theta_j
\end{pmatrix}
-K
\begin{pmatrix}
1 & -c_1 \\
c_1 & 1
\end{pmatrix}
\sum_kL^H_{jk}
\begin{pmatrix}
\rho_k \\
\theta_k
\end{pmatrix}\, .
\end{equation}
Remark that, even if we are perturbing around a limit cycle, namely a time dependent solution, the coefficients of the linearised equations do not depend on time, owing to the specific structure of the GL equation. This observation will simplify the successive analysis, which will follow closely that discussed in the preceding section for the case of a Turing instability. In the next section we will instead deal with a  problem for which the linearised dynamics yields a time dependent Jacobian.

To proceed further we expand the perturbations $\rho_j$ and $\theta_j$ on the Laplacian eigenvectors basis 
\begin{equation}
\begin{pmatrix}
\rho_j \\
\theta_j
\end{pmatrix}
=\sum_{\alpha=1}^{n}
\begin{pmatrix}
\rho_{\alpha} \\
\theta_{\alpha}
\end{pmatrix}
 \phi^{\alpha}_j
\, ,
\end{equation}
inserting the latter into~\eqref{eq:vareqGL}, and by using the orthonormality of the eigenvectors,  we obtain:
\begin{equation}
\label{eq:vareqGLalpha}
\frac{d}{dt}
\begin{pmatrix}
\rho_\alpha \\
\theta_\alpha
\end{pmatrix}
=
\begin{pmatrix}
-2 & 0 \\
-2c_2 & 0
\end{pmatrix}
\begin{pmatrix}
\rho_\alpha \\
\theta_\alpha
\end{pmatrix}
-K\Lambda^H_\alpha
\begin{pmatrix}
1 & -c_1 \\
c_1 & 1
\end{pmatrix}
\begin{pmatrix}
\rho_\alpha \\
\theta_\alpha
\end{pmatrix}\, .
\end{equation}
We put forward the ansatz of exponential growth for each mode, that is $\rho_\alpha \sim e^{\lambda_\alpha t}$ and $\theta_\alpha \sim e^{\lambda_\alpha t}$ and we eventually obtain a condition formally equivalent to the dispersion relation
\begin{equation}
\label{eq:disp_rel}
\lambda(\Lambda^{\alpha})=-(1+K\Lambda^{\alpha})+\sqrt{(1+K\Lambda^{\alpha})^2-K\Lambda^\alpha\left[2(c_1c_2+1)+(1+c_1^2)K\Lambda^\alpha\right]}\, ,
\end{equation}

Let us observe that $\lambda(\Lambda^{1})=0$, signifying that the reference orbit is a limit cycle and thus neutral stable. On the other hand if $\Re \lambda (\Lambda^{\alpha})$ is positive for some $\alpha>1$, the perturbation grows exponentially in time, and the initial homogeneous state proves unstable. Conversely, if $\Re \lambda (\Lambda^{\alpha})< 0$, for every $\alpha$, the perturbation fades away and the system converges back to the fully synchronised state. Expanding~\eqref{eq:disp_rel} for small $K\Lambda^\alpha$ we get
\begin{equation*}
\lambda(\Lambda^{\alpha}) \sim -K\Lambda^{\alpha} (1+c_1c_2)+\dots\, .
\end{equation*}
By recalling that $\lambda(0)=0$, $K>0$ and $\Lambda^{\alpha}>0$ for $\alpha>1$, one can conclude~\cite{CBBCDPF} that $\lambda(\Lambda^{\alpha})>0$ for some $\alpha$ if and only if $1+c_1c_2<0$, that is a necessary and sufficient condition for the loss of stability of the fully synchronised solution.

\begin{figure}[ht]
\centering
\includegraphics[scale=0.35]{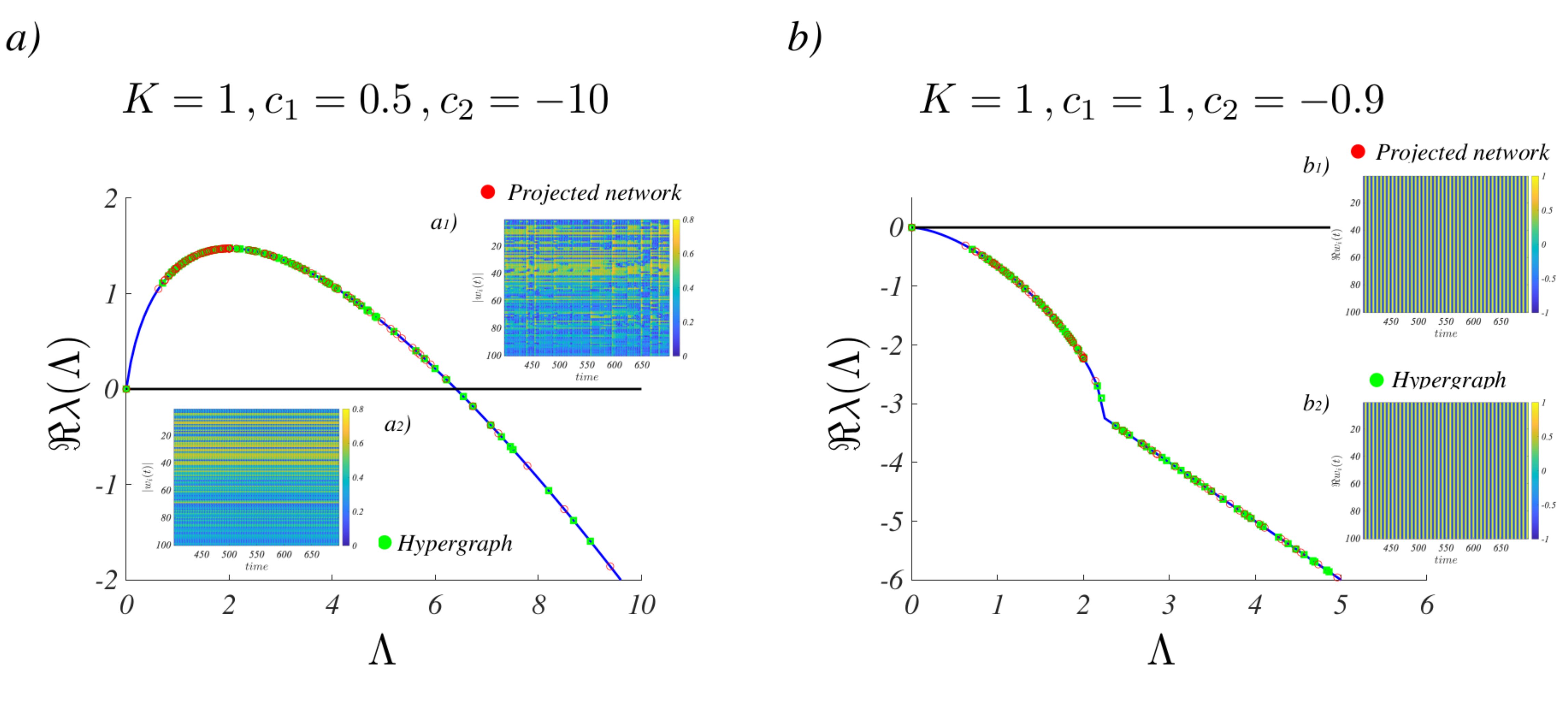}
\caption{\textbf{Synchronisation for Stuart-Landau system}. Main panels: The \textbf{dispersion relation} for the Stuart-Landau system defined on the hypergraph and the projected network is shown for two sets of parameters. In  panel (a) ($K=1$, $c_1=0.5$ and $c_2=10$) this choice yields to a loss of synchronisation. Indeed there are eigenvalues associated to positive values of the dispersion relation and the resulting patterns are heterogeneous (see insets $(a_1)$ for the projected network and $(a_2)$ for the hypergraph). In both insets nodes are ordered for increasing hyper degree, resp. degree, for hypergraph, resp. projected network. The localisation is stronger when the dynamics is hosted on  the hypergraph. In panel (b) the chosen parameters ($K=1$, $c_1=1$ and $c_2=-0.9$) result in the emergence of a globally synchronised state, the dispersion relation being always negative. This can be appreciated by looking at the insets ($(b_1)$ for the projected network and $(b_2)$ for the hypergraph) where we report $\Re W_j$ as function of time.}
\label{fig:reldispGL}
\end{figure}

The numerical results reported in Fig.~\ref{fig:reldispGL} complement the analytical theory discussed above. In  panel (a) of Fig.~\ref{fig:reldispGL} we present the dispersion relation and the heterogeneous patterns emerging for both the hypergraph and the associated projected network, for $K=1$, $c_1=0.5$ and $c_2=-10$. The dispersion relation is positive over a finite domain and the patterns (represented by $\Re W_j(t)$) that develop as follow the instability are pretty localised. In panel (b) of Fig.~\ref{fig:reldispGL}, the parameters are set to the values $K=1$, $c_1=1$ and $c_2=-0.9$. The dispersion relation is non positive and the system displays synchronised oscillations: the imposed perturbation dies out and the oscillators evolve at unison.

\subsection{Master Stability Function on hypergraphs}
\label{ssec:MSF}

In the previous section we have analysed the synchronisation  of an ensemble made of Stuart-Landau (SL) oscillators defined on a hypergraphs. To this end we employed  a straightforward generalisation of the techniques presented in section~\ref{ssec:TP}, when investigating the emergence of Turing patterns. The use of the dispersion relation has been made possible because, for coupled SL equations, the variational problem yields a time independent Jacobian, once evaluated on the periodic homogeneous solution~\eqref{eq:vareqGL}. This is not true for generic nonlinear oscillators. To overcome this problem one can however resort to the formalism of the Master Stability Function (MSF)~\cite{Pecora}, as introduced above. The aim of this section is thus to study the MSF in its full generality for systems defined on hypergraph. In particular, we will set to analyse the synchronisation of nonlinear chaotic oscillators coupled through a hypergraph and compare the outcome of the analysis to that obtained when operating the system on the corresponding projected network.

Let us consider again  Eq.~\eqref{eq:GLHGlinalpha} and replace now in the latter equation $\varepsilon \Lambda_\alpha$ by a generic parameter $\kappa>0$ and thus also the projection $\delta\mathbf{y}_\alpha$ by a generic ``perturbation'' vector $\delta\mathbf{y}$ 
\begin{equation}
\label{eq:GLHGlinalphak}
\frac{d\delta\mathbf{y}}{dt}=\left[D\mathbf{F}(\mathbf{s})-\kappa D\mathbf{G}(\mathbf{s})\right]\delta\mathbf{y}\, .
\end{equation}
The largest Lyapunov exponent of Eq.~\eqref{eq:GLHGlinalphak} is called the Master Stability Function~\cite{Pecora}. Let us denote it by $\lambda(\kappa)$ to emphasise its dependence on the parameter $\kappa>0$. If for all $\kappa$, $\lambda(\kappa)<0$, then $\delta\mathbf{y}$ decays to $0$. At variance, if there exists $\kappa>0$ such that $\lambda(\kappa)>0$, then $\delta\mathbf{y}$ will grow. Back to Eq.~\eqref{eq:GLHGlinalpha} one can conclude that if for a given $\varepsilon$ there exists $\alpha$ such that $\lambda(\varepsilon\Lambda_\alpha)>0$, then the associated $\delta\mathbf{y}_\alpha$  grows in time. Thus individual units deviate from the reference solution $\mathbf{s}(t)$. On the other hand if for all $\alpha$ one has $\lambda(\varepsilon\Lambda_\alpha)<0$ then the system reaches a globally synchronised state: all units will follow at the unison the same chaotic orbit. Let us observe that $\lambda(0)>0$ being the reference orbit, $\mathbf{s}(t)$, a chaotic one. 

To {proceed in the} analysis we assume linear coupling functions~\cite{HCLP}: in this way the MSF simplifies, since $D\mathbf{G}$ is a constant matrix. Moreover, we will assume the matrix $D\mathbf{G}$ to have only one non zero element, say $DG_{ba}$ which denotes the existence of a coupling between the $a$--th and the $b$-th {component of $\mathbf{x}$}.

Let us observe that the variational equation still contains explicitly the time variable via the Jacobian of the reaction part, $D\mathbf{F}(\mathbf{s}(t))$, which is indeed evaluated on the chaotic orbit. Hence to compute the MSF we have to solve a non autonomous system of ODEs, to study the evolution of the norm of $\delta\mathbf{y}(t)$ and then use the definition of the maximum Lyapunov exponent $\lambda(\kappa)=\lim_{t\rightarrow\infty} \frac{\log ||\delta\mathbf{y}(t)||}{t}$. This can result in a tricky exercise. Indeed if $\lambda(\kappa)>0$ then the norm can quickly increase to produce an overflow. On the other hand, if $\lambda(\kappa)<0$, then $||\delta\mathbf{y}(t)||$ shrinks below round-off error. For this reason we employed in our analysis the {\em Mean Exponential Growth of Nearby Orbits} (MEGNO) algorithm~\cite{Megno2,Megno}. This is an improved chaos indicator {that allows to rapidly discriminate} between chaotic and regular orbits. {The method makes it possible for the} Lyapunov exponent {to be consequently recovered}. For these reasons, MEGNO has been largely used in the framework of planetary systems~\cite{GBM2001,LHC2011}, satellites and spatial debris~\cite{VDLC2009,CLD2012,HLDC2013} and also generic nonlinear dynamical systems~\cite{Megno}. The method overcomes the above mentioned limitation by performing a sort of time average of the norm of the deviation vector (see Appendix~\ref{sec:megno}).

Without loss of generality we will use the Lorenz model~\cite{Lorenz1963} for {a demonstrative application}:
\begin{equation}
\label{eq:lorenz}
\begin{cases}
 \dot{x}&=\sigma (y-x)\\
 \dot{y}&=x(\rho-z)-y\\
 \dot{z}&=xy-\beta z\quad .
\end{cases}
\end{equation}
In the following we will fix the model parameters to the ``standard values'', $\beta=2$, $\sigma=10$ and $\rho=28$ {for which the system exhibits the chaotic orbit with a ``butterfly shape''}. Once we couple the above ODE using high-order interactions, i.e. the hypergraph, we get
\begin{equation}
\label{eq:lorenzhg}
\frac{d}{dt}\left(
\begin{smallmatrix}
 x_i\\y_i\\z_i
\end{smallmatrix}
\right)=
\left(
\begin{smallmatrix}
 \sigma(y_i-x_i)\\x_i(\rho-z_i)-y_i\\x_iy_i-\beta z_i
\end{smallmatrix}
\right)
-\sigma \sum_{j} {L}^H_{ij} \mathbf{E}\left(
\begin{smallmatrix}
 x_j\\y_j\\z_j
\end{smallmatrix}
\right)\, ,
\end{equation}
where the constant $3\times 3$ matrix $\mathbf{E}$ encodes for the coupling among the three variables and {its entries take values $0$ or $1$}. For instance if $E_{21}=1$ {and otherwise $E_{ij}=0$}, (noted for short $1\rightarrow 2$) then the growth rate of the second variable, $y$, depends on the first one, $x$, that is $\dot{y}_i\sim -\varepsilon \sum_jL_{ij}^H x_j$ (discarding the reaction part). 

\begin{figure}[ht]
\centering
\includegraphics[scale=0.5]{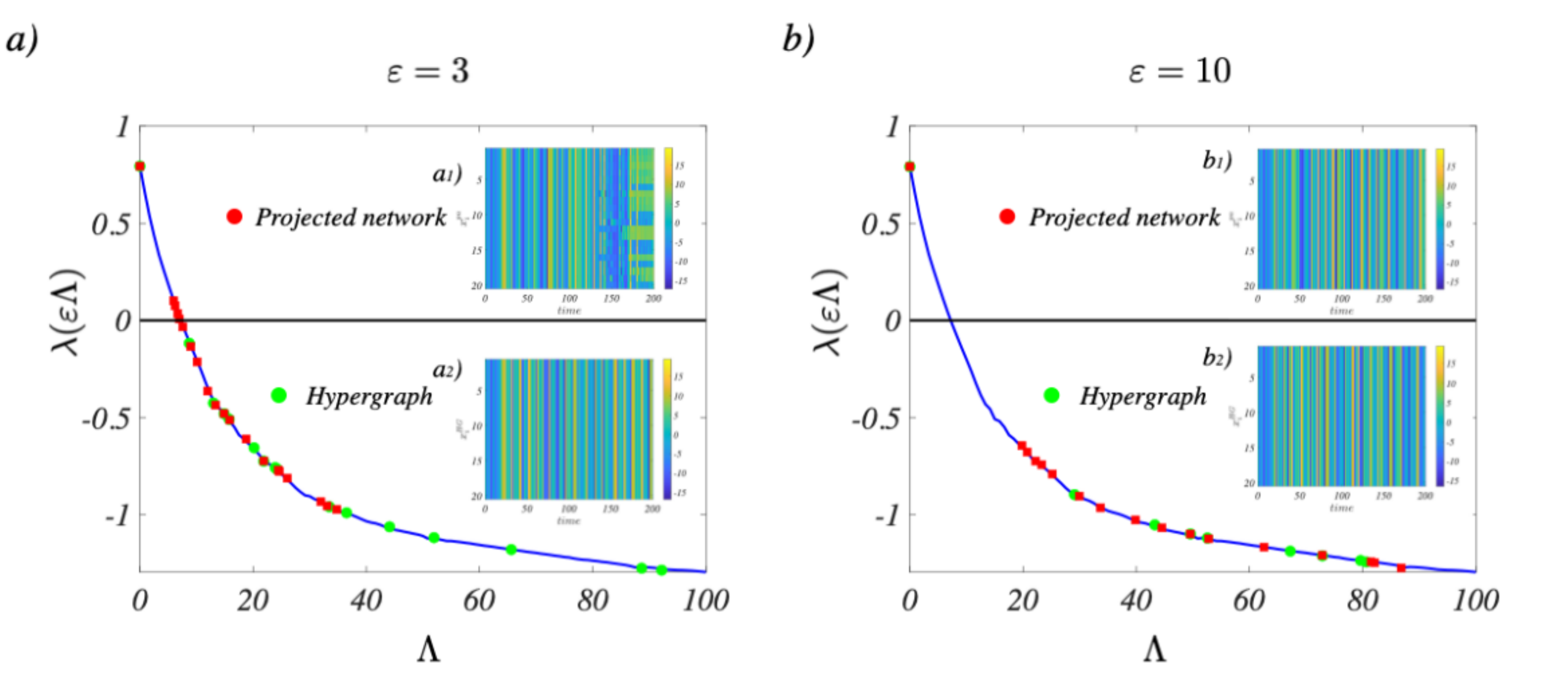}
\caption{\textbf{Master Stability Function and synchronisation for the Lorenz system I}. We report the MSF for the Lorenz model for linear couplings, $1\rightarrow 1$ (main panels) and two choices of the coupling strengths $\varepsilon=3$ (a panel) and $\varepsilon=10$ (b panel). For a small coupling strength (panel (a)), the MSF is negative in correspondence of  the eigenvalues of the Laplace matrix defined  on the hypergraph (green dots), while the MSF can assume positive values once evaluated on the spectrum of the Laplace matrix for the projected network (red dots). In the former case ,the system  synchronises (see inset $(a_2)$) while in the latter it does not (see inset ($a_1$)). For larger coupling strengths (panel (b)) the MSF is negative for both the hypergraph and the projected network and thus, in both cases, the systems do synchronise (see insets $(b_1)$ and $(b_2)$).}
\label{fig:MSF11Lorenz}
\end{figure}

We are now in a position to {adapt the above described theory, i.e. linearise about the reference orbit and project the perturbation on the eigenbase of the Laplace matrix,} to Eq.~\eqref{eq:GLHGlinalpha} {for the case of the Lorenz system.} We will in particular compute the MSF to check the stability of the homogeneous states obtained by replicating chaotic Lorenz trajectory on each node of the collection. In the main panel of Fig.~\ref{fig:MSF11Lorenz} we report the MSF for the coupling scheme, $1\rightarrow 1$, that in the classification proposed in~\cite{HCLP}, corresponds to {class} $\Gamma_1$, {namely the MSF is monotone decreasing and it has a single root.} We consider in particular two values of the coupling strength $\varepsilon=3$ (panel (a)) and $\varepsilon=10$ (panel (b)).  For (sufficiently) small coupling strength (panel (a)), the MSF evaluated on the discrete spectrum of the hypergraph Laplace matrix (green dots) is always negative and thus the system synchronises to the chaotic reference orbit, as shown in the inset $(a_2)$. On the other hand the MSF for the projected network (red dots) takes positive values:  the chaotic oscillators cannot synchronise, as we can appreciate from inspection of inset $(a_1)$. For large enough coupling strength (panel (b)), both spectra yeld a negative MSF (green and red dots in panel (b)) and hence, in both cases, the systems do synchronise (see insets $(b_1)$ and $(b_2)$).

From these results one can draw a first conclusion. Once we fix the coupling strength $\varepsilon$, the sign of the MSF depends on the spectrum of the Laplace matrix for the hypergraph, $\mathbf{L}^H$. {Similarly} for the projected network. However, as we observed in Section~\ref{sec:loceigv} the {eigenvalues of the hypergraph Laplacian extend over a large portion of the real axis, as compared to what it happens when considering the projected network.} Hence the coupling scheme $1\rightarrow 1$ favours the synchronisation on the hypergraph, {provided} the coupling strength is sufficiently small. Said figuratively, one can act on the ``knob'' $\varepsilon$ and have the spectra to slide on the MSF: by progressively reducing the value of $\varepsilon$ one can {force} the spectrum of the projected network to enter  the zone where the MSF is positive, whereas for the same value of $\varepsilon$ the spectrum of the hypergraph is still associated to a negative MSF.

\begin{figure}[ht]
\centering
\includegraphics[scale=0.4]{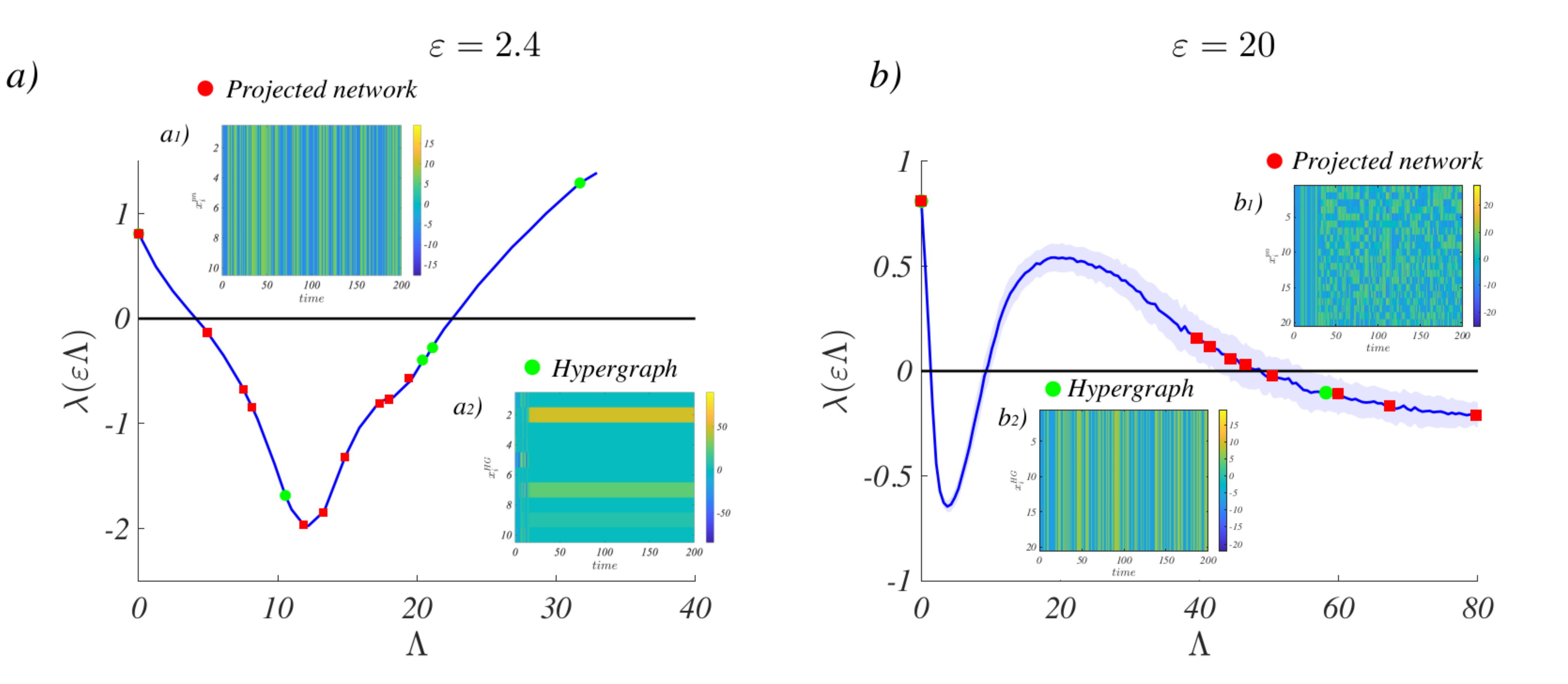}
\caption{\textbf{Master Stability Function and patterns for the Lorenz system II}. We report the MSF for the Lorenz model using the linear couplings, $1\rightarrow 2$ (a panel) and $3\rightarrow 3$ (b panel). In the former case we can observe that, for the chosen value of the coupling strength, $\varepsilon=2.4$, the projected network yields a negative MSF (red dots) and thus the systems synchronises (inset $a_1$), Conversely, the hypergraph possesses unstable eigenmodes (green dots) and the system goes consequently unstable (inset $a_2$). The opposite behaviour is displayed in the case of the $3\rightarrow 3$ coupling for  $\varepsilon=20$ (panel (b)): here the projected network exhibits unstable eigenmodes (red dots) while the hypergraph shows a negative MSF (green dots). The inset $(b_1)$ testifies on the absence of synchronisation for the projected network, while in the inset $(b_2)$ synchronisation is shown to occur on the hypergraph. Calculating the MSF proves less stable for the setting analysed in panel (b). The blue solid line refers to the average computed over $200$ independent runs of the MEGNO algorithm. The shaded region (light blue) refer to the associated standard deviation.}
\label{fig:MSF12e33Lorenz}
\end{figure}

In Fig.~\ref{fig:MSF12e33Lorenz} we report a similar analysis for the coupling schemes $1\rightarrow 2$ (a panel) and $3\rightarrow 3$ (panel (b)). In the classification proposed in~\cite{HCLP} the former corresponds the class $\Gamma_2$, two zeros, while the latter to $\Gamma_3$, three zeros. From the results shown in Fig.~\ref{fig:MSF12e33Lorenz}, one can conclude that the system behaves similarly for couplings $3\rightarrow 3$ and $1\rightarrow 1$:  if the coupling is sufficiently large (here $\varepsilon=20$), synchronisation {is found} on the hypergraph but not on the corresponding projected network. This generalises our previous observation to all couplings belonging to an odd class $\Gamma_{2m+1}$.

The reported behaviour is reversed once we consider couplings that belong to an even class. As we can appreciate from inspection of Fig.~\ref{fig:MSF12e33Lorenz} panel (a) one can choose a sufficiently small coupling to have the MSF negative on the projected network (red dots), while it {takes} positive {values, when} the problem formulated on the hypergraph (green dots). 

\section{Conclusion}
\label{sec:concl}

In this work we {took} a step forward {in modelling dynamical systems on networks. The aim of the work is to account for} high-order interactions {among coupled units}. In particular we focused on the hypergraphs, a very versatile setting where to model systems endowed with many-body interactions. Indeed one can easily represent such high-order interactions via the hyperedge, {so as to overcome the limitations intrinsic to dealing with binary exchanges}.

Starting from a microscopic process {which takes place} on the hypergraph, i.e. a random walk biases toward the size and the number of hyperedges a node belongs to, we defined a new combinatorial Laplace operator {which generalises the concept of diffusive interaction to a multidimensional setting}. This operator reduces to the standard combinatorial Laplacian once the hypergraph {converges back to an ordinary} network. {In this respect, the newly introduced Laplacian can be rationalised as a natural extension of the usual operator}.

In this framework we considered dynamical systems defined on top of hypergraphs and analysed {the stability of the associated homogeneous equilibria}. In particular we extended the Master Stability Function to this {formalism and investigated the specificity of Turing patterns for the generalised proxy of reaction-diffusion systems on hypergraphs. We also analysed the synchronisation of periodic and chaotic orbits, shedding light on the role exerted by high-order couplings.}

In all the inspected cases, the spectral properties of the novel Laplace operator {are central in shaping the ensuing patterns, which appear remarkably localised, as illustrated with reference to the Turing setting.} {Further,} hypergraphs can enhance or impede the synchronisation, {as compared to what it happens on the corresponding} projected network, depending {on the specificity of the imposed couplings.}

\appendix
\section{Compute the MSF using MEGNO}
\label{sec:megno}

To compute the MSF one has to solve {Eq.~\eqref{eq:GLHGlinalpha}}. By discarding the partitioning into reaction and coupling parts, one can rewrite the previous equation as
\begin{equation*}
\frac{d\delta\mathbf{x}_i}{dt}=\sum_{j}\mathcal{J}_{ij}(t)\delta\mathbf{x}_j\, ,
\end{equation*}
that is a time dependent ODE, often named {\em variational equation}. The latter should thus be solved together with the evolution of the reference trajectory
\begin{equation*}
\frac{d\mathbf{x}_i}{dt}=\mathcal{F}(\mathbf{x}_i)\, ,
\end{equation*}
where again we combine in  $\mathcal{F}$ the reaction and the coupling parts.

Then calling $\delta \mathbf{x}(t)=\delta \mathbf{x}(t;\delta \mathbf{x}_0)$ the
solution of the variational equation with initial datum
$\delta \mathbf{x}(0)=\delta \mathbf{x}_0$, the {\em Mean Exponential Growth factor by Nearby Orbits} (MEGNO)~\cite{Megno2,Megno}, can be defined as:
\begin{equation}
  \label{eq:megno}
  Y_{\mathbf{s}}(t):=\frac{2}{t}\int_0^t \frac{\dot{\delta}(\tau)}{\delta(\tau)}\tau\,
  d\tau\, ,
\end{equation}
where $[\delta(\tau)]^2=||\delta \mathbf{x}(\tau)||^2=(\delta \mathbf{x}(\tau),\delta \mathbf{x}(\tau))$, i.e. the norm of the vector $\delta \mathbf{x}$, being $(\cdot,\cdot)$ the scalar product. We also emphasised that the MEGNO is being computed with respect to the reference orbit $\mathbf{s}(t)$. A trivial computation gives:
\begin{eqnarray}
  \label{eq:tricomp}
\frac{d}{dt}\delta^2 &=&2\delta \dot{\delta}\notag\\
&=& (\frac{d}{dt}\delta \mathbf{x}, \delta \mathbf{x})+(\delta \mathbf{x}, \frac{d}{dt} \delta \mathbf{x} )= (\mathcal{J}\delta \mathbf{x}, \delta \mathbf{x})+(\delta \mathbf{x}, \mathcal{J} \delta \mathbf{x} )\, ,
\end{eqnarray}
hence
\begin{equation}
  \label{eq:tricomp2}
  \frac{\dot{\delta}(s)}{\delta(s)}=\frac{(\mathcal{H} \delta \mathbf{x} ,\delta \mathbf{x})}{\delta^2}\, ,
\end{equation}
where $\mathcal{H}=(\mathcal{J}^T+\mathcal{J})/2$ is the Hermitian part of $\mathcal{J}$.

Together with the MEGNO one usually defines also the {\em (time)--averaged MEGNO}:
\begin{equation}
  \label{eq:avermegno}
  \bar{Y}_{\mathbf{s}}(t):=\frac{1}{t}\int_0^t Y_{\mathbf{s}}(\tau)\, d\tau\, .
\end{equation}
$Y(t)$ could in principle display large oscillations for large $t$, so limiting its effective predictive power. At variance, it can be shown that the average-MEGNO is well behaved and allows to study the dynamics for long times. Indeed the main feature of  the average-MEGNO (and/or the MEGNO) is to allow to
distinguish between regular orbits, for which $\bar{Y(t)}\rightarrow 0$, and irregular orbits, for which $\bar{Y}(t)$ grows unbounded. More precisely, $\bar{Y}(t)\sim
\lambda t/2$ where $\lambda$ is the largest Lyapunov characteristic number (or
maximal Lyapunov exponent) of the orbit $\mathbf{s}(t)$. Let us observe that for regular orbits, MEGNO is able to differentiate between periodic ones, ${Y}(t)\rightarrow 0$, and quasi-periodic ones, ${Y}(t)\rightarrow 2$.

Let us observe that one can overcome the problem of the growth of $\delta$ in case of chaotic orbits by employing the following trick. Assume $\delta \mathbf{x}$ to represent a solution of the variational equation. Then one can introduce the ``reduced vector'' $\mathbf{w}$, $\mathbf{w}=\delta \mathbf{x}/\delta$, whose evolution is given by:
\begin{equation*}
  \dot{\mathbf{w}}=\mathcal{J}\mathbf{w}-(\mathcal{H} \mathbf{w},
  \mathbf{w})\mathbf\, .
\end{equation*}
It can easily proven that $||\mathbf{w}(t)||=1$. Indeed 
\begin{eqnarray*}
\frac{d}{dt} ||\mathbf{w}||^2&=& (\frac{d}{dt}\mathbf{w},\mathbf{w})+(\mathbf{w},\frac{d}{dt}\mathbf{w})\\
&=& (\mathcal{J}\mathbf{w},\mathbf{w})-(\mathcal{H}\mathbf{w},\mathbf{w})||\mathbf{w}||^2+(\mathbf{w},\mathcal{J}\mathbf{w})-(\mathcal{H}\mathbf{w},\mathbf{w})||\mathbf{w}||^2\\
&=&2
(\mathcal{H}\mathbf{w},\mathbf{w})(1-||\mathbf{w}||^2)=0\, ,
\end{eqnarray*}
where use has been made of the fact that $||\mathbf{w}(0)||=1$.

\bibliography{bib_HRW}
\end{document}